\documentstyle[twocolumn,prl,aps,epsfig]{revtex}
\newcommand{\beq}{\begin{eqnarray}}
\newcommand{\eeq}{\end{eqnarray}}
\renewcommand{\vec}[1]{{\mathbf{#1}}}
\begin{document}
\draft
%\preprint{dvi file made on \today}

\title
{Transport Properties near the z=2 Insulator-Superconductor Transition}
\author{Denis Dalidovich and Philip Phillips}\vspace{.05in}

%
%\begin{instit}
\address
{Loomis Laboratory of Physics\\
University of Illinois at Urbana-Champaign\\
1100 W.Green St., Urbana, IL, 61801-3080}

   %\end{instit}
%

\address{\mbox{ }}
\address{\parbox{14.5cm}{\rm \mbox{ }\mbox{ }
We consider here the fluctuation conductivity near the point of the
insulator-superconductor transition in a system of Josephson junction
arrays in the presence of particle-hole asymmetry and Ohmic dissipation.
The transition
is characterized by the dynamical critical exponent $z=2$, opening the
possibility of the perturbative renormalization-group (RG) treatment.
The coupling to the Ohmic heat bath, giving the finite quasiparticle
life-time, leads to the non-monotonic behavior of the dc conductivity
as a function of temperature in the leading logarithmic approximation.
}}
\address{\mbox{ }}
\address{\mbox{ }}

%\begin{\multicols}{2}
%\twocolumn
%\columnseprule 0pt \narrowtext
\maketitle

\subsection{Introduction}
In granular superconducting thin films
and fabricated 2D Josephson junction arrays (JJA), phase coherence
is destroyed at zero temperature, whenever the intergrain Josephson tunneling
of Cooper pairs becomes smaller than the Coulomb interaction between grains.
\cite{review,goldman,van der Zant,doniach,otterlo}. The occurrence of
this insulator-superconductor quantum phase transition (IST) can be best
seen from conductivity measurements.  At the transition point, the
resistivity is observed to be temperature-independent and close in value
to the quantum of resistance for charge $2e$ particles
$R_Q=\sigma_Q^{-1}=h/4e^2$. There are many theoretical works devoted
to the explanation of this remarkable result
\cite{otterlo,fisher,wen,cha,fazio,herbut,damle}.
Cha et.al., for instance, calculated the zero-temperature collisionless
conductivity at the transition point within the effective 3D Ginzburg-Landau
(GL) action for an $M$-component bosonic field\cite{cha},
that can be derived from the commensurate 2D Bose-Hubbard model.
The influence of the quartic term in this approach, considered up
to the first order in $1/M$,
was shown to reduce the mean-field value of the conductivity
$(\pi /8)\sigma_Q=\pi e^2/2h$ by 36\%. Fazio and Zappala
calculated this conductivity applying the dimensional regularization
in $\epsilon=3-d$\cite{fazio}. However, as was shown
later by Damle and Sachdev\cite{damle}, the collisionless zero-temperature
conductivity is not the conductivity that is measured experimentally at a
finite temperature. This conclusion stems from the non-commutativity of the
limits $\omega \rightarrow 0, T=0$ and $\omega =0, T\rightarrow 0$ in the
general expression for $\sigma (\omega ,T)$ at the
transition point. The former limit describes collisionless transport,
while it is the latter limit that must be taken to
assess the experimentally-relevant collision-dominated conductivity.
With collisions neglected, the finite temperature dc
conductivity is singular on the insulating side, so the inclusion of
quasiparticle damping is necessary for its regularization.
Collisions are properly accounted for utilizing the quantum kinetic
equation which represents, in general, a complicated non-linear integral
equation for the distribution function of thermally excited quasiparticles.
The kernel of this equation is determined by the quartic non-linearity
in the GL action responsible for collisions between quasiparticles. The
approximate methods, such as $1/M$ or $\epsilon=3-d$ expansion, should be
used to solve the kinetic equation close to the transition point
\cite{damle,sachd1}. Another way of regularization of the singular Drude
conductivity is the inclusion of the phenomenological Ohmic dissipation.
Experimentally, the latter plays a central role, because in
combination with disorder it can lead to the temperature-independent
resistivity as $T\rightarrow 0$\cite{mason}. From the theoretical point
of view Ohmic dissipation is always a relevant perturbation, affecting
thus the functional form of universal quantities near the IST
point\cite{otterlo,denis}. The mentioned sources of dissipation lead
to different temperature dependences of the conductivity, and,
generally speaking, should be treated simultaneously\cite{denis2}.

Thus far, most of the approaches have used as a starting point
the particle-hole symmetric GL action that is space-time isotropic at
$T=0$ and belongs to the universality class with the mean-field dynamical
exponent $z=1$.  However, the behavior in
the critical region can be considerably affected by a term
that breaks the symmetry between particles and holes.
The physical picture leading to the modified action is realized in
a system of JJA in the presence
of uniformly-distributed frustrating offset charges $q_x \neq 2e$
that cannot be eliminated by Cooper pair tunneling\cite{otterlo,grignani}.
The microscopic quantum phase model describing this situation is also
known to be equivalent to the clean incommensurate Bose-Hubbard
model\cite{fisherweich}. The IST in this case is characterized by the $z=2$
dynamical critical exponent. Thus, the quartic term in the GL
action is marginally irrelevant, and there exists the range of parameters
$T$ and $\delta$ (the deviation from the zero-temperature critical point)
that permits all interesting physical quantities
to be calculated in the perturbative renormalization-group (RG) treatment.
\cite{dsfisher,millis,sachsent}.
In this region the static quantities can be regarded in the leading
approximation as some universal functions
of $\delta$ and $T$, independent of the quartic interaction strength, $u$.
This region persists as long as $\delta$ and $T$ are logarithmically
or even double logarithmically small.

In this paper we use this approach to calculate the dc conductivity close
to the point of the IST, governed by $z=2$, within leading
logarithmic accuracy. It is shown, using the Kubo formula, that in the
absence of any dissipative mechanism, the real part of finite temperature
conductivity is singular and non-universal on the insulating side.
Its regularization by inclusion of
the coupling to the Ohmic heat bath leads to the logarithmic
temperature dependence in the quantum critical (QC) regime.
The conductivity  increases in the leading approximation
as $\ln\ln(1/T)$ with $T\rightarrow 0$ right above the transition,
and decreases monotonically in the quantum disordered (QD) regime.
Consequently, we claim then that crossing over from the QD to the QC
regime with decreasing temperature is accompanied by a somewhat re-entrant
behavior of the dc conductivity.
The regularization that results from the finite lifetime of
quasiparticles, is possible
only once the Umklapp scattering is included. This is a consequence of
the Galilean invariance of the action, which explicitly
describes the system of interacting
charges of one sign. These processes ensure that the inverse lifetime of
quasiparticles, $1/\tau_U \sim e^{- A/T}$, leads, in the absence of
other sources of dissipation, to an exponentially large
fluctuation conductivity on the insulating side.
We discuss the applicability
of this approach to the description of the finite-$T$ charge transport
near the 2D metal-superconductor transitions.

\subsection{Renormalization Group Analysis}

The general form of the GL functional that models the behavior near
the IST point in the presence of uniformly-distributed offset charges is
derived to be\cite{otterlo}
\beq\label{action}
F[\psi]&=&\int d^2r\int d\tau\left\{
\left[\left(\nabla+\frac{ie^*}{\hbar}\vec A(\vec r,\tau)\right)
\psi^*(\vec r,\tau)\right]\right.\nonumber\\
&&\left.\cdot
\left[\left(\nabla-\frac{ie^*}{\hbar}\vec
A(\vec r,\tau)\right)\psi(\vec r, \tau) \right] +
\lambda \psi^{\ast}(\vec r,\tau) \partial_\tau\psi(\vec r,\tau)
\right.\nonumber\\
&&\left.+ \kappa^2\left|\partial_\tau\psi(\vec r,\tau)\right|^2
+\delta \left|\psi(\vec r,\tau)\right|^2 +
\frac{u}{2}\left|\psi(\vec r,\tau)\right|^4 \right\} \nonumber\\
&& +F_{\rm dis}
\eeq
where $\vec{A} (\vec r,\tau)$ is the vector potential, due to applied
electric field, $e^*=2e$, and  $\delta$ is proportional to the inverse
correlation length. In Fourier space, the dissipation term,
$F_{\rm dis}=\eta \sum_{\vec k, \omega_n} |\omega_n||\psi(\vec k, \omega_n)|^2$
corresponds to the Ohmic model of Caldeira and Leggett\cite{caldeira}.
Parameters $\kappa$ and $\lambda$ (linear
in the value of offset charges\cite{otterlo}) measure the strength
of quantum fluctuations. We will regard $\kappa /\lambda =O(1)$.
The term with $\kappa$ is clearly irrelevant in the region of interest,
so we can neglect it and measure all quantities, having the
dimensionality of energy, in units
of $\lambda$. The key idea behind the calculation of the general,
frequency-dependent conductivity in the critical region is that in 2D
at finite temperature, the conductivity obeys the scaling\cite{sondhi,zwang}
\beq\label{scalecond}
\sigma_{\alpha\beta}(\delta ,T,\omega ,u)=
\sigma_{\alpha\beta}(T(l^*),\omega(l^*),u(l^*)).
\eeq
In the momentum-shell RG, $l^*$ denotes the scale at which the
effective size of the Kadanoff cell is on the order of the
correlation length.  We assume without loss of generality that the
scaling stops when $\delta (l^*)=1$. The frequency in the above
equation scales trivially, $\omega (l)=\omega e^{zl}$, although
we will be interested only in the dc conductivity, most commonly
measured in experiments. The dissipation term with $\eta$, preserving the
$z=2$ universality class, should be also incorporated in the RG equations.
However, because any non-zero $\eta$ profoundly changes the
analytic structure of a quasiparticle propagator, it is convenient to
consider the cases of zero and non-zero $\eta$ separately.

{\bf $\eta=0$:}
The one-loop finite-$T$ RG equations for this problem in 2D are of the
form\cite{dsfisher}
\beq\label{T}
\frac{dT(l)}{dl}=zT(l)
\eeq

\beq\label{delta}
\frac{d\delta (l)}{dl}=2\delta (l)+\frac{2K_2u(l)}{\exp [(1+\delta (l))/T(l)]}
\eeq

\beq\label{u}
\frac{du(l)}{dl}=(2-z)u(l) - C[\delta(l),T(l)] u(l)^2,
\eeq

where the coefficient
\beq \label{C}  C[\delta(l),T(l)] \approx
\left\{ \begin{array}{ll}
        \displaystyle\frac{K_2}{2(1+\delta (l))} & \quad T(l) \ll 1 \\[4mm]
        \displaystyle\frac{5K_2 T(l)}{(1+\delta (l))^2} & \quad T(l) \gg 1
        \end{array}\right.
\eeq
Here $K_2=(1/2\pi)$, and in the one-loop approximation we can assume $z=2$.
The solution of the above system is qualitatively different
in the quantum disordered and quantum critical regimes.

{\it Quantum disordered regime:}
We are in the QD regime, if the inequality $\delta \gg T$ between
the bare parameters holds.
With sufficient accuracy the term with $u(l)$ in the righthand side of
Eq. (\ref{delta}) can be neglected, and we obtain
in this regime, that $l^* =\frac{1}{2} \ln(1/\delta )$, giving
\beq\label{qdtu}
T^*=\frac{T}{\delta}, \quad
u^* \approx \frac{8\pi }{\ln\frac{1}{\delta}}.
\eeq
The asterisk denotes, that the parameters under scaling
must be taken at $l=l^*$.
The result for $u^*$ is obtained with logarithmic accuracy, i.e
we consider $l^*$ to be large. This condition ensures the validity of the
perturbation theory.

{\it Quantum critical regime:}
This region is characterized by the opposite condition, $\delta \ll T$.
The integration of the RG equations in this case consists of two steps.
In the first step, we integrate over $l$ from $0$ to $\tilde{l}$, such that
$T(\tilde{l})=1$. We have then
$\delta (\tilde{l})=(\delta /T +O(1/\ln T^{-1})) \ll 1$.
So, the scaling must continue into the second step in which $T(l)\gg 1$.
In this, classical region, the strength of interactions is measured
not by $u(l)$, but by $v(l)=u(l)T(l)$\cite{millis}, and for $l>\tilde{l}$
we must consider the equations
\beq\label{v}
\frac{dv(l)}{dl}=2v(l)-\frac{5K_2 v(l)^2}{(1+\delta (l))^2}
\eeq

\beq\label{newdelta}
\frac{d\delta (l)}{dl}=2\delta (l)+\frac{2K_2 v(l)}{1+\delta (l)}
\eeq
Assuming, that throughout the scaling $v(l)\ll 1$, that is we are working in
the weakly classical region, we obtain, using $v(\tilde{l})=4\pi /\tilde{l}$,
that $v(l)\approx (4\pi/\tilde{l}) e^{2(l-\tilde{l})}$.
Substitution of this result into Eq. (\ref{newdelta}) and subsequent
integration gives us with the logarithmic accuracy the equation for
$l^*$ in the QC regime:
\beq\label{eqlstar}
\delta(l^*)=\frac{4}{\tilde{l}} (l^* -\tilde{l}) e^{2(l^*-\tilde{l})}=1.
\eeq
Solving this within double logarithmic accuracy, we find that
\beq\label{qclstar}
l^*=\frac{1}{2} \ln \left[ \frac{1}{T} \ln\ln (\frac{1}{T}) \right],
\eeq

\beq\label{qdtv}
T^*=\ln\ln \frac{1}{T}, \quad v^*=\frac{2\pi}{\ln\ln\frac{1}{T}}
\eeq
The validity of the above solution requires not only
the condition $\ln(1/T) \gg 1$, but also $\ln\ln(1/T) \gg 1$.

{\bf $\eta \ne 0$:} The presence of the term $\eta |\omega_n|$ in the
RG equation for $\delta$ leads to the divergency in the sum over frequencies.
So the momentum-shell RG with the upper frequency cutoff $\Gamma$ should be
used\cite{millis}, and the obtained one-loop equations are of the form
\beq\label{delta2}
\frac{d\delta (l)}{dl}=2\delta (l)+f^{(2)}(\delta(l),T(l)) u(l)
\eeq

\beq\label{u2}
\frac{du(l)}{dl}=-f^{(4)}(\delta(l),T(l)) u(l)^2,
\eeq
where $f^{(2)}$ and $f^{(4)}$ are some rather robust functions
of $\delta(l)$, $T(l)$ and $\eta$, that we don't write out here. It is
essential, that the results for $T^*$ and $l^*$ remain the same in both
the QC and QD regimes, if we are interested only in the leading logarithmic
approximation. However, the subdominant temperature dependent corrections
appear to be  different. $u^*$ and $v^*$ are also logarithmically small,
though have the cutoff dependent numerical prefactors.

\subsection{Transport Properties}
We see, that upon scaling, for small enough $T$ and $\delta$, the parameters
$u^*$ and $v^*$ remain small, and in the first approximation one can
calculate the conductivity with the help of the Kubo formula,
\beq
\sigma_{\alpha\beta}(i\omega_n)=-\frac{\hbar}{\omega_n}\int d^2r\int d\tau
\frac{\delta^2\ln Z}{\delta \vec{A}_\alpha(\tau,\bf r)\delta \vec{A}_\beta(0)}
e^{i\omega_n\tau},\nonumber
\eeq
applied to the Gaussian part of Eq.(\ref{action}). The standard calculations
for the longitudinal conductivity, which we denote simply as $\sigma$,
lead to the result
\beq\label{cond}
\sigma(i\omega_n^*)&=&\frac{2(e^*)^2}{\hbar\omega_n^*}T^*\sum_{\omega_n^*}
\int \frac{d^2k}{(2\pi)^2}2k_{x}^2 G(\vec k,\omega_m^*)\nonumber\\
&&\left(G(\vec k, \omega_m^*)-G(\vec k,\omega_m^*+\omega_n^*)\right),
\eeq
where $G(\vec k,\omega_n^*)=(i\omega_n^*+\epsilon_k ^*)^{-1}$ is the usual
Matsubara Green function. The rescaled temperature $T^*$ and the energy
of quasiparticles $\epsilon_k^*=1+k^2$ are employed in the righthand side
in accordance with Eq. (\ref{scalecond})
($\omega_m^*=2\pi mT^*$). We must perform then the analytical continuation to
real frequencies after doing the summation over frequencies $\omega_m^*$.

If the absence of any dissipative mechanism, we obtain:
\beq
\sigma (\omega^*)=\left[ \frac{i}{\omega^*}+\pi \delta (\omega^*)
\right] \frac{(e^*)^2}{2hT^*} \int_0^\infty
\frac{k^{3}dk}{\sinh^2 \left( \displaystyle\frac{1+k^2}{2T^*} \right) }.
\eeq
Though the integral over $k$ is calculable exactly, we state here only the
results in the limiting cases of the QD and QC regimes:
\beq\label{collesscond}
\sigma(\omega )=\left( \frac{i}{\omega }+\pi \delta (\omega ) \right)
\frac{(e^*)^2}{h} T
\left\{ \begin{array}{ll}
              e^{-\delta /T} & \quad {\rm QD} \\
         \ln\ln\ln \frac{1}{T} & \quad {\rm QC}
        \end{array}\right.
\eeq
The expression in the QC regime is written with triple
logarithmic accuracy. We see that on the insulating side, the real
part of the collisionless conductivity does not have any regular contribution,
as $\omega \rightarrow 0$.
The singular part disappears as well at $T=0$.
Moreover, the form of the frequency dependence in Eq. (\ref{collesscond})
is characteristic for the conductivity of a superconducting phase.
This unphysical result suggests
that the finite life-time of quasiparticles should be accounted for in the
correct determination of transport properties.

The inclusion of the non-zero Ohmic dissipation provided by the last term in
Eq. (\ref{action}), changes the analytical properties of Matsubara sums,
and the analytical continuation in Eq. (\ref{cond}) should be
performed by contour integration
\beq\label{condcont}
\sigma(\omega^*)&=&\frac{(e^*)^2}{h\omega^*}
\int_0^\infty \!\! k^3 dk \int_{-\infty}^{\infty}
\!\!\coth\frac{z}{2T^*} dz\left[(G^R(z)-G^A(z))\right.\nonumber\\
&&\left.[G^R(z)+G^A(z)-G^R(z+\omega^*)-G^A(z-\omega^*)]\right],
\eeq
where the retarded and advanced Green functions
$ G^{R/A}(z)=(z+\epsilon_k^* \mp i\eta z)^{-1}$ have been introduced.
In the limit $\omega \rightarrow 0$, expansion of the corresponding Green
functions and subsequent integration by parts yields the dc conductivity:
\beq\label{dccon}
\sigma &=&\frac{(e^*)^2}{2\pi h} \int_0^\infty k^3
dk \int_{-\infty}^{\infty}\frac{dx}{\sinh^2 x}\nonumber\\
&&\times\frac{8(\eta T^* x)^2}{\left[ (\epsilon_k^*
+2x T^*)^2+(2\eta T^* x)^2 \right]^2}.
\eeq

In the QD regime $T^* =T/\delta \ll 1$, and, as long as $\eta \ll T/\delta$,
the integral over $x$ is determined by the two competing contributions:
from the region near the minimum of denominator at
$x_0=-\epsilon_k^*/2T^*$ and from the vicinity of $x=0$. Evaluating
those contributions and integrating subsequently over momentum $k$,
we receive the total conductivity:
\beq\label{qdcond}
\sigma=\frac{2e^2}{h} \left[ \frac{T}{\delta \eta} e^{-\delta /T}
+\frac{2\pi}{9} \left( \frac{\eta T}{\delta}\right)^2 \right].
\eeq
As can be seen, the first term is dominant for very small values of
dissipation, $\eta < (\delta/T)^{1/3}e^{-\delta/3T}$, while for larger
$\eta$ only the second contribution should be retained. In any
of that cases the conductivity monotonically decreases as $T\rightarrow 0$.

In the QC regime the double logarithmic accuracy implies that
$T^* = \ln\ln(1/T) \gg 1$. With the same accuracy, the main contribution
to the integral over $x$ comes from $x \ll 1$, allowing us to set
$x^2/\sinh^2 x \approx 1$. We obtain then after simple integrations,
assuming $\eta \leq 1$, that
\beq\label{qccond}
\sigma=\frac{e^2}{h} \frac{1+\eta^2}{\eta} \ln\ln\frac{1}{T}.
\eeq
This suggests, that the conductivity is not universal in the QC regime
and increases upon lowering the temperature, albeit very slowly, as a double
logarithm of $1/T$. One can conclude that,
because of the crossover from the QC to the QD region upon
lowering temperature for any non-zero $\delta$,
it may be possible to observe a non-monotonic behavior of the resistivity
as a function of temperature with a dip when $T \sim \delta$.
Though the derived results are strictly applicable in rather tiny
regimes, where $\delta$ and $T$ are logarithmically small, the results
found here may be observable beyond those boundaries.

Now consider the role of mutual scattering of quasiparticles as the other
possible source of dissipation. Eq.(\ref{action}) suggests that our
action is Galilean invariant, describing the system of charge carries of
one definite sign. In the Hamiltonian
formalism, our system is identical to that of weakly-interacting Bose
particles in 2D\cite{lifshitz}. The linearization of the corresponding
collision integral in the kinetic equation in a small correction to the
equilibrium distribution, proportional to ($\vec k \vec E$),
\beq\label{kinetic}
\left( \frac{\partial}{\partial t} + e^* \vec E(t) \frac{\partial}
{\partial \vec k} \right) f(\vec k, t) = I[f(\vec k,t)],
\eeq
reveals, that $I[f(\vec k,t)]$ turns to zero as a consequence of
momentum conservation.
This situation is different from the $z=1$ case, in which the system
consists of colliding particles and holes having the same energy\cite{damle}.
This does not mean, however, that the mutual collisions of quasiparticles in
the $z=2$ case can not lead to the finite lifetime. The appropriate
way to see this is to go beyond the continuum limit and recall that the
real experimental systems, such as JJA, have a lattice
periodicity. This suggests that
in case of a square lattice, the quasiparticle energies
$\epsilon_{\vec k}=4-2(\cos k_x +\cos k_y)$
are periodic functions of the quasimomentum $\vec k$,
and the Umklapp processes
need to be taken into account. The corresponding inverse scattering time
$1/\tau_U$ can be estimated from\cite{lifshitz}
\beq\label{tauumklapp}
\frac{1}{\tau_U} \propto u^2 \int n(\epsilon_{\vec k_1})
                          (1+n(\epsilon_{\vec k_2}))
\delta(\epsilon_{\vec k}+ \epsilon_{\vec k_1}-
     \epsilon_{\vec k_2}- \epsilon_{\vec k_3}) \nonumber\\
\cdot (2\pi)^2 \delta (\vec k+ \vec k_1 -\vec k_2 -\vec k_3-\vec b)
\frac{d^2k_1}{(2\pi)^2}\frac{d^2k_2}{(2\pi)^2}\frac{d^2k_3}{(2\pi)^2},
\eeq
where the integration runs over the first Brillouin zone, and $\vec b$ is the
reciprocal lattice vector. The non-zero $\vec b$ implies that the conditions
imposed by $\delta$-functions on quasimomenta and energies of
colliding quasiparticles can both be satisfied only if one of the
incoming and one of the outgoing quasimomenta come from the
interior of the first Brillouin zone. Because the main contribution, as can
be seen from the Kubo formula, comes from small $k$,
$k_2$ must be large. Physically, this means that the damping of the
critical fluctuations of the order parameter arises from the collisions with
the non-critical modes. One can easily estimate
that $1/\tau_U \propto e^{-A/T}$
where the prefactor varies slowly $\delta$ and $T$, and
$A$ is some numerical constant determined by the concrete form
of the first Brillouin zone and the quasiparticle dispersion law.
This inverse scattering rate, valid in both QC and QD regimes, is
very small near the quantum phase transition point. This implies that
in the absence of any Ohmic dissipation $\eta$ the conductivity
$\sigma \propto \tau_U \propto e^{A/T}$, and hence exponentially increases
with temperature. The latter conclusion can be verified by making the
substitution $\eta xT \rightarrow (1/2\tau_U)$ in Eq. (\ref{dccon}).
Such behavior suggests that Ohmic dissipation plays the dominant role near
the 2D IST point in the presence of charge frustration and reinstates
the insulating behavior.
For JJA, fabricated from s-wave superconductors, weak non-magnetic disorder
is non-pair breaking. Hence, as long as the pairs remain intact,
it can not lead to dissipative terms in our action\cite{anderson}.
Ohmic dissipation arises if the superconducting grains
are embedded in a conducting environment\cite{wag} in which case
Eqs. (\ref{qdcond}) and (\ref{qccond}) give the conductivity due to
fluctuations of the superconducting order parameter.

Conventional disorder leads, however, to dissipation in
a d-wave superconductor because in this case it is a pair-breaking
perturbation\cite{herbut2}. The finite temperature fluctuation
conductivity near this disorder-driven metal-superconductor
quantum phase transition can be investigated using the same action
(\ref{action}), provided we are not too close to the transition point
where the disorder affects the critical properties.
Depending on parameters of the microscopic Hamiltonian, $\eta$ can be both
small and large, compared to unity\cite{herbut2}.
If $\eta \leq 1$, the conductivity in the QC regime is given
by Eq. (\ref{qccond}), and we should use the last term of Eq. (\ref{qdcond})
for the QD regime. If $\eta \gg 1$, the relevant conductivity
was discussed using $N=\infty$ approach\cite{denis}.

This work was funded by the DMR98-12422 of the NSF research fund.

\end{document}